\documentclass[runningheads]{llncs}
\usepackage{eccv}
\usepackage{eccvabbrv}  

\usepackage{graphicx}
\usepackage{booktabs}
\usepackage{soul}
\usepackage{subcaption}

\usepackage[accsupp]{axessibility}  
\usepackage{hyperref}
\usepackage{orcidlink}

\begin{document}

\title{Lung tumor segmentation in MRI mice scans using 3D nnU-Net with minimum annotations} 

\titlerunning{Lung Tumor Segmentation in MRI Mice Scans}

\author{Piotr Kaniewski, Fariba Yousefi,
   Yeman Brhane Hagos, \\Talha Qaiser, Nikolay Burlutskiy\\
}


\institute{AstraZeneca R\&D UK}

\maketitle

\begin{abstract}
    In drug discovery, accurate lung tumor segmentation is an important step for assessing tumor size and its progression using \textit{in-vivo} imaging such as MRI. While deep learning models have been developed to automate this process, the focus has predominantly been on human subjects, neglecting the pivotal role of animal models in pre-clinical drug development. In this work, we focus on optimizing lung tumor segmentation in mice. First, we demonstrate that the nnU-Net model outperforms the U-Net, U-Net3+, and DeepMeta models. Most importantly, we achieve better results with nnU-Net 3D models than 2D models, indicating the importance of spatial context for segmentation tasks in MRI mice scans. This study demonstrates the importance of 3D input over 2D input images for lung tumor segmentation in MRI scans. Finally, we outperform the prior state-of-the-art approach that involves the combined segmentation of lungs and tumors within the lungs. Our work achieves comparable results using only lung tumor annotations requiring fewer annotations, saving time and annotation efforts. This work\footnote{\url{https://anonymous.4open.science/r/lung-tumour-mice-mri-64BB}} is an important step in automating pre-clinical animal studies to quantify the efficacy of experimental drugs, particularly in assessing tumor changes.
    
\end{abstract}

\section{Introduction}
\label{section:intro}

Lung cancer is the leading cause of global cancer incidence and mortality, and it is characterized by the growth of abnormal cells in the tissues of lungs~\cite{lung_cancer_epidemiology, tumor_growth1, tumor_growth2}. The detection of those abnormal cells has been revolutionized by deep learning which accelerated and facilitated the process of detecting, classifying, and annotating tumors, often with human-like accuracy~\cite{dl_revolution}. Although advanced models like the segment anything model (SAM), transformers, or stable diffusion are being designed and applied at the intersection of oncology and medical imaging for cancer patients~\cite{sam, transformer_survey, mamba, medsegdiff}, increasingly greater focus in research is also being put on animals. 

Animal models like mice or rats are essential in the pre-clinical pipeline of drug discovery and development due to their biological similarities to human genetic and physiological characteristics. It provides a critical platform for metabolic, safety, and efficacy studies~\cite{mouse_metabolic}, allowing one to develop a full understanding of a drug and disease~\cite{mouse_disease} and translate it onto humans~\cite{mouse_to_human}. Despite inherent limitations and ethical considerations, the use of animal models is a regulatory requirement, and corroborating drugs selected for clinical trials have a solid basis for pre-clinical validation. The conventional methods for quantifying tumors in the pre-clinical domain are laborious, time-intensive, and prone to observer variability.    

Magnetic resonance imaging (MRI) is a type of biological imaging method, a non-invasive method to render images of living tissues. Although the segmentation of tumors within small animals' organs is on the rise, majority of research is focusing on brain tumor segmentation with the use of MRI datasets~\cite{brain1, brain2, brain3}. While MRI scans have also been used for cardiac strain segmentation in rats~\cite{cardiac1, cardiac2}, the lung tumor segmentation in mice focuses only on utilizing CT scans~\cite{mouse_ct1, mouse_ct2, yolo_tumor} which involve potentially harmful X-ray radiation~\cite{ct_radiation}. Although CT scans are the usual imaging technique for lung tumor detection, tumors can be also detected with the help of MRI which is non-invasive and can be safely applied to children or pregnant women regularly, allowing for tracking of the tumor growth over time~\cite{deepmeta}. Furthermore, such 2D and 3D MRI segmentation of lung tumor was successfully achieved in human patients~\cite{human_lung_mri}, highlighting a potential gap in \textit{in-vivo} animal image segmentation.

One potential reason for lack of research in the field of pre-clinical lung tumor segmentation using MRI scans could be the lack of data. To date, there is only one publicly available MRI dataset\footnote{\url{https://zenodo.org/records/7014776}} with lungs and tumors annotated, acquired by~\cite{deepmeta} who introduced a method DeepMeta which outperformed U-Net on the task of lung and tumor segmentation. Although with the use of this model, the authors were capable of recognizing growth pattern of tumors, the segmentation was conducted for both lungs and tumors, indicating that tumor segmentation could rely heavily on contextual information coming from background organs. While such a model is somewhat viable from clinical and anatomical perspectives, it would be beneficial to also have a \textit{sole-tumor} segmentation model as often clinicians might not be interested in lung segmentation. However, such sole-tumor segmentation would be much more challenging to achieve due to the small sizes of tumors and the contrast being much less clear.

Furthermore, even though the provided MRI dataset contained 3D scans, the DeepMeta study only focused on 2D models, neglecting potentially useful spatial context. This is likely due to GPU constraints as training a model on 3D images can be computationally expensive~\cite{computational_limit}. 
Although tumors can be successfully segmented using 2D images, there is evidence that 3D segmentation improves the overall performance for isotropic datasets~\cite{nnunet_isotropic}. Therefore, although 2D models can be computationally cheaper, it would be highly desirable to have an architecture that is capable of effectively segmenting lung tumors in both 2D and 3D settings. While 3D models can leverage the spatial advantages of 3D imaging, they are computationally more expensive. In contrast, 2D models offer greater flexibility in terms of resource and time availability but lack the spatial context.

One such architecture could be "no-new-Net" (nnU-Net), a deep learning based segmentation method that can automatically configure the network architecture, training environment, and pre-processing stage~\cite{nnunet_nature}. The framework configures itself automatically based on the data fingerprint and set of defined rules which aims for best performance and computationally efficient results. Even though the framework is based on the U-Net architecture, with a few modifications, it outperformed state-of-the-art approaches on numerous occasions, highlighting its simplicity and importance of data-oriented approach~\cite{call_for_validation}. Due to its automatic pre-processing strategy, it enables for 2D and 3D segmentation, allowing to examine the importance of spatial context in \textit{in-vivo} imaging~\cite{thorax}.

In this work, we aim to address the gap in the field of \textit{in-vivo} medical imaging and examine nnU-Net performance on the task of tumor segmentation in 2D and 3D MRI scans of mice. We reproduced the results of the U-Net, U-Net3+, and DeepMeta models, benchmarked them against the nnU-Net framework, and then extended the work to the new task of \textit{sole-tumor} segmentation using both 2D and 3D images. The main contributions of this paper are as follows:

\begin{itemize}
    \itemsep0em 
    \item We demonstrated that 3D models outperformed 2D models when we evaluated their performance using nnU-Net. The results indicate the importance of 3D spatial context for 3D lung tumor segmentation tasks in MRI mice scans.
    \item We trained the nnU-Net model that outperformed the previous state-of-the-art models for tumor segmentation in MRI mice scans including U-Net, U-Net3+ and DeepMeta models.
    \item We trained segmentation models using only tumor annotations whereas the previous study used both lungs and tumor annotations~\cite{deepmeta}. Our models performed on par with the previous ones but required less human annotations. 
\end{itemize}

\section{Data}
\label{section:data}

\begin{figure}
    \centering
    \includegraphics[width=0.95\linewidth]{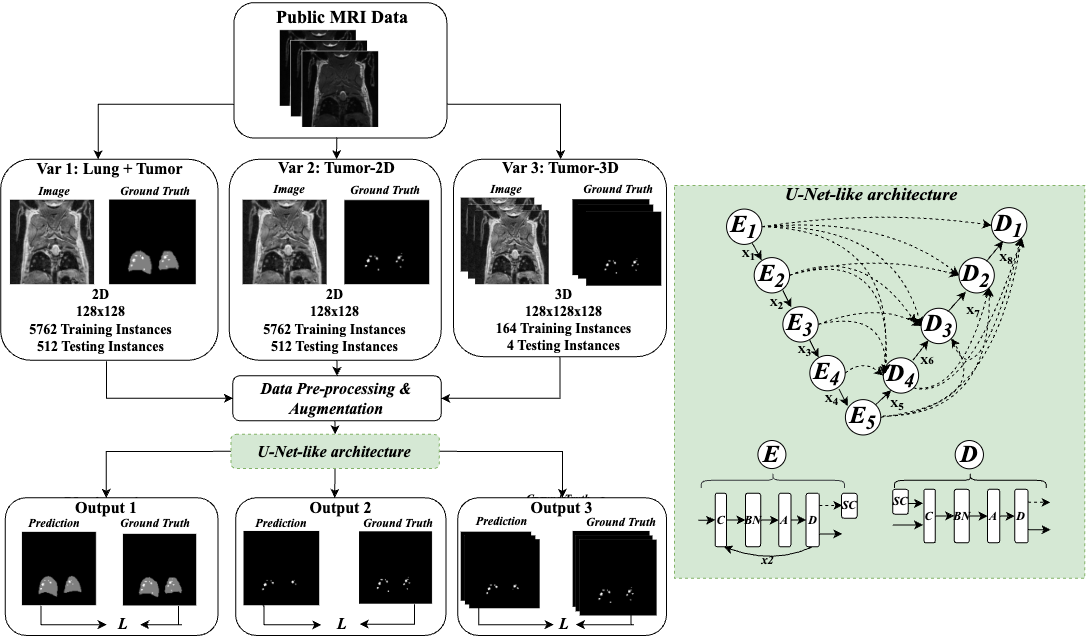}
    \caption{The schematic diagram summarizing our work on lung tumor segmentation. Three different data variants were acquired from public MRI dataset, pre-processed, and used for training one of the four U-Net-like architectures. The architectures differed in terms of loss functions, number of kernels, activation functions, batch-normalization layers, and full-scale skip connections. \textbf{A} - Activation; \textbf{C} - Convolutional Layer; \textbf{D} - Dropout; \textbf{BN} - Batch Normalization; \textbf{SC} - Skip Connections;  \textbf{L} - Loss Function.}
    \label{fig:main_diagram}
\end{figure}

The data used in this study is publicly available~\cite{deepmeta}.
In summary, out of $206$, 3D MRI scans provided, only $168$ had lungs and tumors annotated and described in the associated metadata. Each of those $168$, 3D images had size of $128\times128\times128$ and the annotations were created after each 3D image was sliced into $128$ individual 2D slices. We used $164$, 3D images for training and $4$, 3D images for testing. We utilized the images and associated masks in three different data variants to examine different aspects of MRI lung tumor segmentation using 2D or 3D images. 

\subsection{Data variants}

We examined three data variants when training the models, as mentioned in Figure~\ref{fig:main_diagram}:
\begin{enumerate}
    \item \textbf{Variant 1: \textit{Lung+tumor}} -  consisted of the original 2D images and annotations as used in~\cite{deepmeta}. The 2D slices were extracted from the 3D image,  that contain lung regions. Suppose $(X, Y)$ represents a pair of raw 3D image $X$ and ground truth lung segmentation $Y$, 2D slices which contain lung $S$, are identified using Equation (\ref{eq:lung_selection}).

\begin{equation}\label{eq:lung_selection}
    S = \{(X_i, Y_i) \mid_{i=1}^{n} \sum(Y_i) > 0\} 
\end{equation}

where $X_i$ and $Y_i$ represent the $i^{th}$ slice of a 3D images $X$ and $Y$, respectively. This resulted in  $5762$, 2D slices containing lung tissues. Therefore, lung was always visible in the slices for this variant, while lung tumor was occasionally present. This curated variant was used to explore models' ability to segment lung tumors in a multi-class setting, when organ context is provided in the ground truth masks.
    \item  \textbf{Variant 2: \textit{tumor-2D}} - consisted of the original 2D images by modifying annotations (\textit{i.e.} removing lung labels), leaving just the tumor annotations. Therefore, the $5762$, 2D slices in this variant are identical as in variant 1, however, there is no organ context provided in the ground truth masks. This variant was used to explore models ability to segment only lung tumors in a binary classification setting.
    \item \textbf{Variant 3: \textit{tumor-3D}}  -  consisted of $164$, 3D images with annotations of only lung tumors. Contrary to the remaining variants, here the images contained non-lung tissues as essentially \textit{raw} 3D MRI images were used for the analysis. This variant was used to explore models ability to segment only lung tumors from 3D context.
\end{enumerate}

The provided MRI scans were acquired in two separate batches resulting in different brightness of the scans. To ensure consistent brightness of images, contrast was enhanced for the 'darker' batch of images. The images and masks were normalized to zero mean and unit variance as shown in Equation (\ref{eq:norm}) and then were augmented through rotation and elastic transformation for each data variant. After data augmentation, there were $46096$, 2D scans in the training set for \textit{Lung+tumor} and \textit{tumor-2D} variants, and $1312$, 3D scans in the training set for \textit{tumor-3D} variant. 

\begin{equation}\label{eq:norm}
    X^{'} = \frac{X - \mu}{\sigma}
\end{equation}
where $\mu$ and $\sigma$ are the mean and standard deviation of the input image $X$, respectively and $X^{'}$ is the normalised image. For the sake of simplicity, the $X$ could be a 2D or a 3D image.

There were two variants of tumors present in this dataset, one with small (under $0.4$ mm$^3$) tumors and another with large (over $0.4$ mm$^3$) tumors. Training set contained images with both types of tumors as well as images with no tumors. It is important to note that the images in the training sets of \textit{Lung+tumor} and \textit{tumor-2D} variants always contained lung tissues as slices with no lung tissues were excluded during the data preparation stage. This was not the case for the images in the training set of \textit{tumor-3D} variant, as the entire 3D stacks were used for training. The test set consisted of images of four mice: one without any tumor, two with small tumors, and one with large tumors. For the test sets of \textit{Lung+tumor} and \textit{tumor-2D} variants, each 3D image was sliced into $128$, 2D slices, yielding $512$ images containing both lung and non-lung tissues, contrary to the training set. For the test set of \textit{tumor-3D}, no slicing was done and full 3D images were used for analysis. The contrast was enhanced for three mice coming from 'darker' batch to ensure consistency between train and test sets. 

\section{Our method}
\label{section:methods}
We used the three data variants to examine neural networks architectures' performance on segmenting lung tumor: in a 2D multi-class setting (when both lung and tumor are annotated), in a 2D binary setting (with only tumor being annotated), and in a 3D binary setting. We examined three commonly-used architectures for biomedical segmentation as well as popular nnU-Net, an automatic self-configured framework. We built upon the work by~\cite{deepmeta} and the performance of these architectures was evaluated by assessing their performance not only on segmenting 2D scans, but also on segmenting 3D scans when 3D data was examined.

While all of these architectures were used in the past for the task of biomedical segmentation of human lungs' tumors, they were not used for segmenting tumors using MRI scans of mice except~\cite{deepmeta}, and they did not benchmark the performance against different data variants. The model in~\cite{deepmeta} needs lung segmentation in addition to tumor segmentation. Investigation into sole-tumor segmentation would be beneficial for medical studies from time perspective as it is tumors that are of interest when monitoring cancer progression, not the lungs themselves - therefore annotating the lungs can often be unnecessary and time-consuming.  


\subsection{Deep learning architectures}

In total, four architectures were examined: U-Net~\cite{unet}, U-Net3+~\cite{unet3p}, DeepMeta~\cite{deepmeta}, and nnU-Net~\cite{nnunet_nature}. All of these networks are architecturally similar: they share encoder-decoder framework, bottleneck, double convolutions, and same-scale skip connections. They are composed of five encoding blocks, with the last block often being referred to as bridging block, and four decoding blocks, as represented in Figure~\ref{fig:main_diagram}.

The encoding pathway for original U-Net as well as U-Net3+ starts with $64$ filters which double with each consecutive block. Each block consists of the repeated application of two $3\times3$ un-padded convolutions, followed by activation and down-sampling through max-pooling. The activation function within the blocks was Rectified Linear Unit (ReLU) across all architectures except for nnU-Net which utilized its variant, leaky ReLU. The U-Net, U-Net3+ and DeepMeta architectures contain batch normalization layers within its encoding blocks, and nnU-Net incorporates instance normalization instead. The decoding pathway for U-Net-like architectures involves up-sampling of the feature maps, $2\times2$ up-convolutions which halve the number of filters, and concatenation with respective skip connections. This is followed by repeated $3\times3$ convolutions with ReLU activation functions. The final layer consists of $1\times1$ convolution which maps the logits to respective classes.

Both U-Net3+ and DeepMeta architectures differ from the remaining architectures mainly by the presence of full-scale skip connections. Contrary to U-Net, U-Net3+ and DeepMeta models contain intra-connections between the decoding blocks and full-scale inter-connections between encoding and decoding blocks. This results in a different decoding pathway structure, rather than being symmetrical to the encoding pathway, each decoding block receives feature maps from five different scales, resulting in $320$ and $160$ kernels in each decoding block for U-Net3+ and DeepMeta, respectively. Additionally, the DeepMeta model has standard convolution layer replaced by separable convolution, and its encoder starts with only $16$ filters instead of $64$. All these architectures, except for nnU-Net, are used as 2D networks suitable only for 2D images. 

The exact number of filters remained unchanged for all architectures except for U-Net3+ for which we reduced the initial number of filters from $64$ to $32$ due to high computational cost. 
It is important to note that as nnU-Net framework is automatically self-configuring, therefore its architecture was adapted based on the data fingerprints, which led to the initial number of filters being $32$. 

\subsection{Training}
For training U-Net, U-Net3+ and DeepMeta on \textit{Lung+tumor} or \textit{tumor-2D} variants of the data, we used the same initialization as in~\cite{deepmeta}. Cosine annealing scheduler was used for changing the learning rate. The initial learning rate of $0.001$ was used with batch size of $64$. The slices from the training dataset were split using a ratio of $80:20$ for training and validation, and the entire network was trained for $100$ epochs. 

For nnU-Net, the training environment was configured automatically following the set of fixed rules based on the data fingerprint. The default learning rate of $0.01$ was used for training the 2D models on \textit{Lung+tumor} and \textit{tumor-2D} variants, with poly learning rate policy. The network was trained for $250$ epochs, with batch size of $199$. However, when the framework was trained on 3D images, the initial learning rate was decreased to $0.001$ and the number of epochs was increased to $500$ with batch size of only $2$. Interestingly, the nnU-Net framework did not extract patches from the images and original images were used as input: $128\times128$ for 2D and $128\times128\times128$ for 3D. This is because nnU-Net prioritizes large patch sizes allowing for more contextual information to be aggregated. 

We used Tesla V100-PCIE-32GB GPU for training both 2D and 3D models, and we limited our computational resources to one GPU for fair comparison.  

\subsection{Model parameter optimization functions}


The training dataset can be represented as a set of $N$ pairs, \(\{(X_i, Y_i)\}_{i=1}^{N}\), where \(X_i\) is the $i^{th} $ raw image and \(Y_i\) is the $i^{th}$ corresponding ground truth image for model training. The $X_i$ and $Y_i$ could be either 2D or 3D image depending on the data variant.The model estimates $\hat{Y_i}$ from \(X_i\) as follows: \(\hat{Y_i} = f(X_i)\), where \(f\) is the function learned by the model and \(\hat{Y_i}\) is the predicted segmented image. To optimize model parameters, different loss functions estimating the error between the ground truth \(Y_i\) and the estimated segmented \(\hat{Y_i}\) were applied. 

Majority of these functions were combinations of region-based or distribution-based losses. The only exception was U-Net's loss which only utilized weighted cross entropy as shown in Equation \ref{eq:wce} \cite{unet}. For U-Net3+, a sum of Intersection over Union (IoU) $\mathcal{L}_{\text{IoU}}$ (Equation (\ref{eq:iou}), Focal $\mathcal{L}_{\text{fl}}$~\cite{focal_loss} (Equation \ref{eq:focal}), and Multi-scale Structural Similarity index $\mathcal{L}_{\text{ms-sim}}$~\cite{mssim_loss} (Equation \ref{eq:mssi}) loss functions were used as the compound loss function $\mathcal{L}_{\text{fl}} + \mathcal{L}_{\text{ms-ssim}} + \mathcal{L}_{\text{iou}}$. Focal loss $\mathcal{L}_{\text{fl}}$ was also utilized by DeepMeta model in combination with cross entropy $\mathcal{L}_{ce}$ (Equation \ref{eq:ce}) and Lovasz-Softmax $\mathcal{L}_{lovasz}$~\cite{lovasz-softmax_loss} (Equation \ref{eq:lsl}) losses, yielding \( \alpha \times \mathcal{L}_{\text{ce}} + \beta \times \mathcal{L}_{\text{lovasz}} + \gamma \times \mathcal{L}_{\text{fl}} \) where $\alpha$, $\beta$ and $\gamma$ corresponded to $0.7$, $0.4$, and $0.2$, respectively~\cite{deepmeta}. Cross-entropy $\mathcal{L}_{\text{ce}}$ was also utilized by nnU-Net loss function as it used the sum of cross-entropy and dice loss $\mathcal{L}_{\text{dice}}$ (Equation \ref{eq:dce}) as a loss function: \( \mathcal{L}_{\text{ce}} + \mathcal{L}_{\text{dice}}\)~\cite{nnunet_nature,nnunet_loss}. \\

\begin{equation}\label{eq:wce}
    \mathcal{L}_{\text{wce}} = \sum_{{x} \in \Omega} w({x}) \log({p}_{\ell({x})}({x})),
\end{equation}
where $\ell:\Omega \rightarrow \{1,\dots,K\}$ is the true label of each pixel and $w:\Omega \rightarrow {R}$ is a weight map assigning importance to specific pixels, boundary pixels were assigned higher weight~\cite{unet}. \\

\begin{equation}\label{eq:focal}
    \mathcal{L}_{\text{fl}} = - (1 - p_t)^\gamma \log (p_t),
\end{equation}
where $(1 - p_t)^\gamma$ is a modulating factor to the cross-entropy loss, $\gamma \ge 0$ is a tunable focusing parameter, and $p_t$ is the model's estimated probability for the label class. \\

\begin{equation}\label{eq:iou}
    \mathcal{L}_{\text{IoU}} = 1 - \frac{\sum_i p_i g_i}{\sum_i p_i + \sum_i g_i - \sum_i p_i g_i},
\end{equation}
where $p_i$ represents the predicted value for the $i^{th}$ pixel $g_i$ corresponds to the ground truth value for the $i^{th}$ pixel. \\

\begin{equation}\label{eq:mssi}
    \mathcal{L}_{\text{ms-sim}} = 1 - \prod_{m=1}^{M} (\frac{2\mu_p \mu_g + C_1}{\mu_p^2 + \mu_g^2 + C_1})^{\beta m} \cdot (\frac{2\sigma_{pg} + C_2}{\sigma_p^2 + \sigma_g^2 + C_2})^{\gamma m},
\end{equation}
where \( M \) is the total number of scales, \( \mu_p, \mu_g \) represent mean intensities of \( p \) and \( g \), \( \sigma_c, \sigma_d \) indicate standard deviations of \( p \) and \( g \),\( \sigma_{pg} \) is a covariance between \( p \) and \( g \), while \( \beta_m, \gamma_m \) are defining the relative importance of the two components at each scale. The \( C_1\) and \( C_2\) correspond to $0.01$ and $0.03$, respectively~\cite{mssim_loss}, and they are added to prevent division by zero and stabilize the computation. \\

\begin{equation}\label{eq:ce}
    \mathcal{L}_{\text{ce}} = -\sum_{i=1}^{C} y_i \log(\hat{y}_i), 
\end{equation}
where $y_i$ is the true label of the $i^{th}$ class and $\hat{y}_i$ is the predicted probability of the $i^{th}$  class. \\

\begin{equation}\label{eq:lsl}
    \mathcal{L}_{\text{lovasz}} = \frac{1}{|\mathcal{C}|} \sum_{c \in \mathcal{C}} \Delta_{\mathcal{J}}(\vec{m}(c)),
\end{equation}
where $\mathcal{C}$ is the cardinality, $\vec{m}(c)$ is a vector of pixel errors $m(c)$ for class c, and $\Delta J$ is a Jaccard loss. \\

\begin{equation}\label{eq:dce}
    \mathcal{L}_{\text{dice}} = 1 - \frac{2 \sum_i p_i g_i}{\sum_i p_i^2 + \sum_i g_i^2},
\end{equation}
where similarly to $p_i$ represents the predicted value for the $i^{th}$  pixel and $g_i$ corresponds to the ground truth value for the $i^{th}$ pixel.

\subsection{Performance evaluation}
\label{section:methods:performance}
The models' performance was evaluated using IoU and F1 scores. We evaluated the models' performance by segmenting the test set consisting of four 3D MRI scans ($4 \times 128$, 2D slices stacked along \textit{z} axis). 



\section{Experiments}
Three separate experiments were conducted. Firstly, the models were trained on the \textit{Lung+tumor} variant of the data to predict both lungs and tumors and to examine how well the tumors are detected when both lungs and tumors are annotated. This task included reproducing the work by~\cite{deepmeta} and using nnU-Net on the same data for comparison. In the second experiment, we removed the lung annotations (leading to \textit{tumor-2D} variant of the data), and re-trained the models to segment only tumors. This was done to examine the network performance without the help of underlying lung context. For the last experiment, we used nnU-Net and trained on \textit{tumor-3D} variant of the data for exploring the importance of spatial context when segmenting tumors.

\subsection{Lung and tumor segmentation in 2D setting}
\label{subsec:reproduction_results}

The results obtained after implementing DeepMeta's workflow,
are presented in Table~\ref{tab:reproduce-nopost-table}, in the "Post-processed" columns and they align with the findings reported in the DeepMeta paper. It is important to note that their models' performance was evaluated after the application of a post-processing to the predicted masks. The post-processing stage was a part of the published DeepMeta network pipeline and it involved detecting and removing slices without mouse tissues by application of Laplacian of Gaussian filter~\cite{laplacian, deepmeta} to each slice, and removing blobs smaller than ten pixels for lungs and three pixels for tumors. 


To qualitatively compare the different models, we 1) applied the same post-processing stage to nnU-Net output on the test set for thorough comparison, and 2) compared the 'raw' performance of the trained models on the test sets, with no post-processing being applied. We report those results in Table~\ref{tab:reproduce-nopost-table}.

\begin{figure}
    \centering
    \begin{subfigure}[t]{0.45\linewidth}
        \centering
        \caption*{U-Net}
        \includegraphics[width=0.7\linewidth]{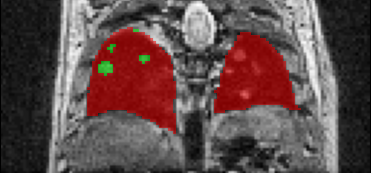} 
        \caption{F1 (lung): 0.95,\\ F1 (tumor): 0.59}
    \end{subfigure}
    ~
    \begin{subfigure}[t]{0.45\linewidth}
        \centering
        \caption*{U-Net3+}
        \includegraphics[width=0.7\linewidth]{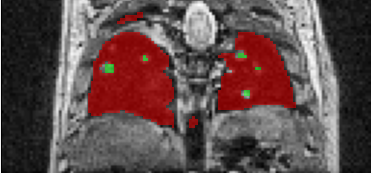} 
        \caption{F1 (lung): 0.92,\\ F1 (tumor): 0.46}
    \end{subfigure}
    ~
    \begin{subfigure}[t]{0.45\linewidth}
        \centering
        \caption*{DeepMeta}
        \includegraphics[width=0.7\linewidth]{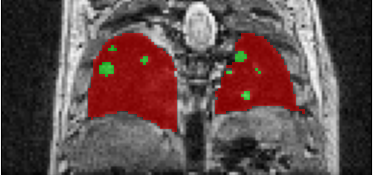} 
        \caption{F1 (lung): 0.95,\\ F1 (tumor): 0.80}
    \end{subfigure}
    ~
    \begin{subfigure}[t]{0.45\linewidth}
        \centering
        \caption*{nnU-Net}
        \includegraphics[width=0.7\linewidth]{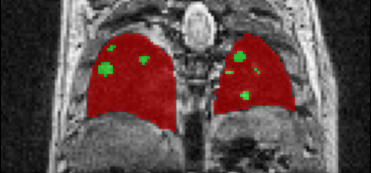} 
        \caption{F1 (lung): 0.96,\\ F1 (tumor): 0.84}
    \end{subfigure}
    ~
    \begin{subfigure}[t]{0.45\linewidth}
        \centering
        \includegraphics[width=0.7\linewidth]{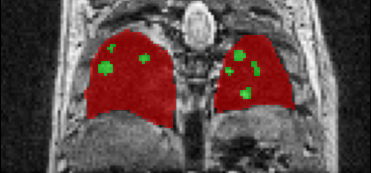} 
        \caption{Ground Truth}
    \end{subfigure}
    \caption{Qualitative results for lung and tumor segmentation. Showing great performance of nnU-Net for both lung and tumor segmentation. Images were obtained without post-processing and were cropped for better visualization purposes.}    
    \label{fig:prediction-masks-lung-tumor}
\end{figure}

The results show that post-processing stage significantly improves the performance of all models, with nnU-Net reaching the best performance for both lung and tumor segmentation. When no post-processing is applied, nnU-Net retains the best performance for tumor prediction however it is outperformed by the remaining models when it comes to lung prediction score. The qualitative analysis of the prediction masks showed that the decline in lung prediction score is due to nnU-Net making false predictions on slices without lung tissues, which significantly lowers the score despite lung tissues being segmented well, as indicated in Figure~\ref{fig:prediction-masks-lung-tumor}. This is due to all training data containing lung tissues, which results in the model always expecting a lung - interestingly, the remaining models made false predictions less often, resulting in better scores for lung detection.

Although DeepMeta model achieved the second-best score when post-processing was applied, it performed worse than U-Net or U-Net3+ when raw predictions were scored. Interestingly, even though DeepMeta design was based on U-Net3+ architecture, U-Net outperformed U-Net3+ regardless of post-processing strategy, indicating that better results could be obtained with simpler architecture. Potentially, the main power of the overall network lays in the fusion loss function combined with post-processing which results in a well segmented lungs and tumors regardless of the architecture. 


\begin{table}
  \caption{Quantitative results for lung and tumor segmentation with and without post-processing (removing non-mouse tissues and small blobs for lungs and tumors) stage on Lung$+$tumor dataset, indicating a comparable performance of nnU-Net with the task of tumor segmentation.}

  \label{tab:reproduce-nopost-table}
  \centering
  \begin{tabular}{llllll}
    \toprule
     & &  \multicolumn{2}{c}{IoU} & \multicolumn{2}{c}{F1}\\
    \cmidrule(r){3-4} \cmidrule(r){5-6}  
    Model & Class  & Post-processed & Raw & Post-processed & Raw \\
    \toprule
 U-Net &  Lungs & $ 0.86\pm 0.03$  & $0.83\pm 0.05 $& $0.92\pm 0.03 $& $0.89\pm 0.06 $\\ 
 & tumors & $ 0.70\pm 0.10$ & $0.67\pm 0.07 $& $0.72\pm 0.10 $& $0.69\pm 0.07 $\\ 
  \midrule
 U-Net3+ & Lungs & $ 0.85\pm 0.04 $&$ \textbf{0.83}\pm \textbf{0.03} $&$ 0.91\pm 0.04 $&$ \textbf{0.89}\pm \textbf{0.03} $\\ 
 & tumors & $ 0.67\pm 0.06 $&$ 0.65\pm 0.05 $&$ 0.69\pm 0.05 $&$ 0.67\pm 0.05 $\\ 
  \midrule
 DeepMeta & Lungs & $ 0.87\pm 0.03 $& $0.80\pm 0.06 $& $0.92\pm 0.04 $& $0.86\pm 0.07 $\\
 & tumors & $ 0.72\pm 0.05 $&$ 0.66\pm 0.04 $&$ 0.75\pm 0.06 $&$ 0.70\pm 0.06 $\\ 
  \midrule
 2D nnU-Net & Lungs &$ \textbf{0.88} \pm \textbf{0.04}$ & $ 0.67\pm 0.03 $&$ \textbf{0.93} \pm \textbf{0.04} $&$ 0.75\pm 0.05 $\\
 & tumors & $ \textbf{0.76}\pm \textbf{0.18} $ &$ \textbf{0.72}\pm \textbf{0.15} $&$ \textbf{0.79} \pm \textbf{0.17}$ &$ \textbf{0.78}\pm \textbf{0.13} $\\
    \bottomrule
  \end{tabular}
\end{table}

\begin{table}[!hbt]
  \caption{Quantitative results for \textit{sole-tumor} segmentation on tumor-2D data variant, with no post-processing applied.}
  \label{tab:reproduce2-table}
  \centering
  \begin{tabular}{lllll}
    \toprule
    Model  & Class & IoU &  F1 \\
    \toprule
 U-Net  & tumor & $ 0.63\pm 0.38$ & $0.64\pm 0.38 $\\ 
 \midrule
 U-Net3+  & tumor & $ 0.70\pm 0.19 $&$ 0.72\pm 0.19 $\\ 
 \midrule
 DeepMeta  & tumor & $ 0.66\pm 0.22 $& $0.66\pm 0.22 $\\
 \midrule
 2D nnU-Net  & tumor & $ \textbf{0.74}\pm \textbf{0.15} $&$ \textbf{0.77}\pm \textbf{0.13} $\\
    \bottomrule
  \end{tabular}
\end{table}

\subsection{Tumor segmentation in 2D setting}
\label{section:only-tumor}

\begin{figure}
    \centering
    \begin{subfigure}[t]{0.45\linewidth}
        \centering
        \caption*{U-Net}
        \includegraphics[width=0.7\linewidth]{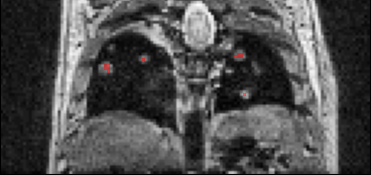} 
        \caption{F1 (tumor): 0.27}
    \end{subfigure}
    ~
    \begin{subfigure}[t]{0.45\linewidth}
        \centering
        \caption*{U-Net3+}
        \includegraphics[width=0.7\linewidth]{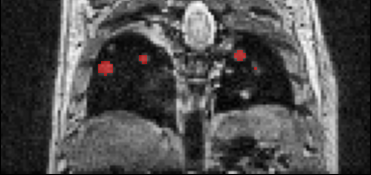} 
        \caption{F1 (tumor): 0.67}
    \end{subfigure}
    ~
    \begin{subfigure}[t]{0.45\linewidth}
        \centering
        \caption*{DeepMeta}
        \includegraphics[width=0.7\linewidth]{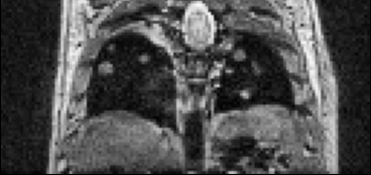} 
        \caption{F1 (tumor): 0.0}
    \end{subfigure}
    ~
    \begin{subfigure}[t]{0.45\linewidth}
        \centering
        \caption*{nnU-Net}
        \includegraphics[width=0.7\linewidth]{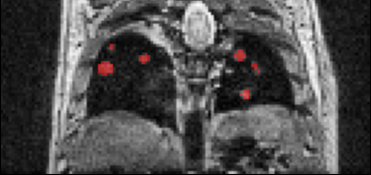} 
        \caption{F1 (tumor): 0.82}
    \end{subfigure}
    ~
    \begin{subfigure}[t]{0.45\linewidth}
        \centering
        \includegraphics[width=0.7\linewidth]{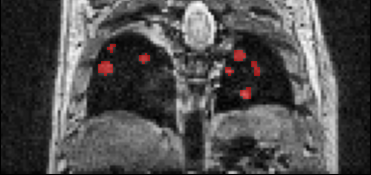} 
        \caption{Ground Truth}
    \end{subfigure}
    \caption{Qualitative results for tumor segmentation, indicating high performance of nnU-Net. Images are cropped for better visualization purposes.}
    \label{fig:prediction-masks-tumor}
\end{figure}

To examine models' performance on the task of segmenting tumors without the help of contextual information, the models were trained on the \textit{tumor-2D} variant of the data, with no post-processing being applied to thoroughly benchmark the models. Some of the qualitative and quantitative results are shown in  Figure~\ref{fig:prediction-masks-tumor} and Table~\ref{tab:reproduce2-table}, respectively. 
 
There is a clear decline in performance of U-Net, U-Net3+, and DeepMeta. Although the mean results only differ by approximately $10\%$, standard deviation is much larger for these models. This is because the models achieve high score on 'healthy' slices (slice without tumors, as they correctly make no predictions) while they fail at segmenting actual tumors, showing low predictive power. 

Such poor performance is likely due to lack of spatial context: when the models were trained on \textit{Lung+tumor} variant of the data, they learnt that tumor masks are spatially related to lung' masks. In \textit{tumor-2D} variant the models need to segment small areas of interest without spatial context, thus the area where tumors are located is not confined by the additional mask. However, this was not an issue for nnU-Net as its performance remained relatively unchanged after re-training on \textit{tumor-2D} variant only.

The qualitative results emphasize the challenge of detecting only lung tumors: while all models segmented tumors relatively well when the lung masks were provided, this was not the case for binary segmentation as only nnU-Net performed well across the whole test set. The remaining models segmented the tumors correctly only occasionally and they also yielded many false positive predictions. Surprisingly, the visual inspection highlighted that DeepMeta has no predictive power and is not segmenting any slices on the provided test set. The relatively high score is due to correctly predicting all slices without tumors which constitute a majority in the test set. Lack of predictive power could be due to specificity of its loss - even though the weight and the functions were modified accordingly, the penalization for false positive prediction is greater than for false negative, therefore the model learns to make no predictions. Although it was found efficient in the multi-class setting, its loss function is task-specific and not transferable to other tasks, unlike the remaining loss functions which resulted in a greater predictive power. This was particularly evident with nnU-Net where the loss function appeared to be highly versatile. 


\subsection{Tumor segmentation in 3D setting}
\label{section:robust_3d}

\begin{figure}
    \centering
    \includegraphics[width=0.8\linewidth]{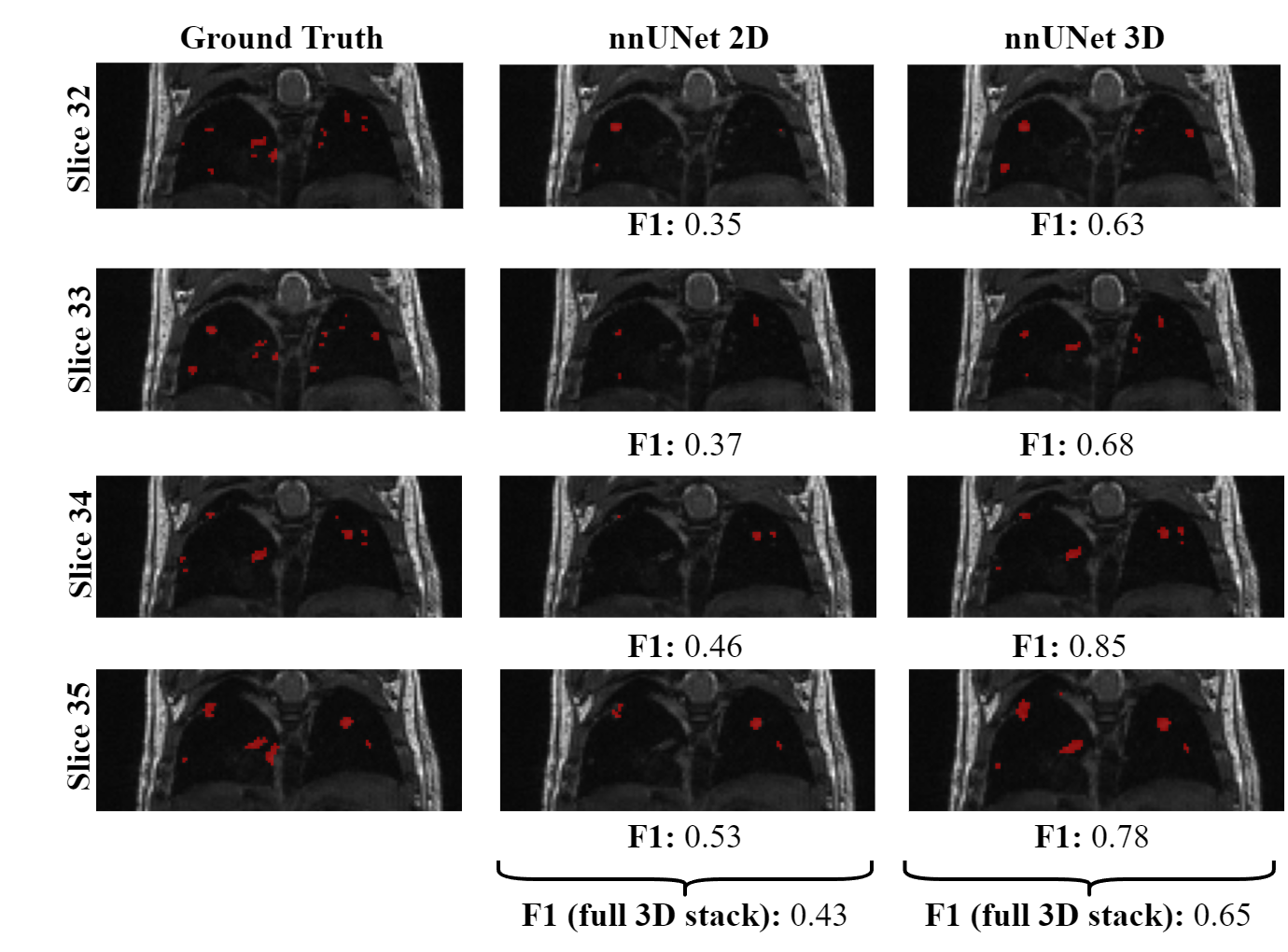}
    \caption{F1-score results on a sequence of slices for 2D and 3D nnU-Nets on a few slices and the full 3D stack of one image. 3D nnU-Net  outperforming other models. }
    \label{fig:nnunet_3d}
\end{figure}

The next step was to examine the models' performance on the \textit{tumor-3D} variant. This task was not only to examine the robustness of the models but also the impact of the spatial context on semantic segmentation. We evaluated both 2D and 3D architectures using the \textit{tumor-3D} variant of the data. This involved utilizing the 3D data as both 2D slices and 3D stacks in order to compare the outcomes between the 2D and 3D models. It's worth noting that in the previous two variants, the slices without lungs were excluded from the analysis which is not the case in this experiment. 

As mentioned in Section~\ref{section:methods:performance}, there are two ways of assessing the performance of semantic segmentation when 3D images are provided. For comparison purposes and consistency, we evaluated the 3D models by using both stack-by-stack and slice-by-slice approaches. The results can be found in Table~\ref{tab:nnunet-table}.

The results indicate clear benefit of using 3D architectures over 2D architectures when analyzing 3D images. The 2D nnU-Net showed slight decline in performance and increase in standard deviation when training on \textit{tumor-3D} variant of the data.

The 2D nnU-Net also performed poorly when assessing the segmentation masks of whole 3D images, however this is to be expected as the architecture and pre-processing strategy are adjusted for segmenting 2D slices.

The results improved when 3D nnU-Net was trained on the \textit{tumor-3D} variant - not only did it achieve higher score, segmenting 3D images as a whole but also better score when assessing 2D slices individually, following slice-by-slice basis. This highlights the importance of spatial context when analyzing the MRI scans. 

The importance of spatial context is further highlighted in Figure~\ref{fig:nnunet_3d} which compares differences between 2D and 3D models. Once spatial context is provided, the tumors are captured more extensively and reflect ground truth better.
Nevertheless, these results indicate a clear benefit of using 3D architectures over 2D architectures when analysing 3D images, even if one is only interested in examining the segmentation masks on slice-by-slice basis.

\begin{table}
  \centering
  \caption{Quantitative results for segmenting lung tumor in 2D and 3D, using \textit{tumor-3D} data variant.}
  \label{tab:nnunet-table}
  \begin{tabular}{lllll}
    \toprule
     & IoU & F1 \\
     \midrule
 nnU-Net 2D model  &$0.69 \pm 0.20$ & $0.71 \pm 0.20$ \\  
    \midrule
 nnU-Net 3D model  & $\textbf{0.73 } \pm \textbf{0.19 }$ & $\textbf{0.75 } \pm \textbf{0.18}$ \\  
    \bottomrule
  \end{tabular}
\end{table}

\subsection{Conclusion and future work}
\label{section:conclusion}

Our experiments demonstrated that the nnU-Net outperformed the previous state-of-the-art models including DeepMeta, U-Net, and U-Net3+ for lung tumor segmentation in MRI mice scans. Notably, nnU-Net framework is flexible for developing models on 3D images without many manual architectural adjustments. Also, the nnU-Net data-oriented framework offers a robust solution that can be easily tuned for a different number of classes, such as lung and tumor classes, and for various spatial resolutions. 

Most importantly, our study demonstrated the importance of training on 3D images instead of training on 2D images, slice-by-slice basis, for segmentation tasks in MRI mice scans. Training models using 3D images provide important spatial context that is crucial for 3D segmentation tasks.

Finally, to our knowledge, our work is first-of-its-kind on semantic segmentation of lung tumors in MRI mice scans using solely 3D tumor masks, achieving comparable performance to prior study, where both lung and tumor masks were used which is more time and effort consuming. 

As a future direction, we envision that active learning will help to identify and improve the limitations of such models. Annotating images is resource-intensive, and implementing an active learning loop to identify the most informative images will enable one to retrain the model with minimal data, mitigating covariance shift issues and enhancing efficiency. Finally, we believe that there will be more research work on exploring 3D versus 2D modeling using biological imaging.


%
%
\bibliographystyle{splncs04}
\bibliography{ref}

\newpage

\appendix
\setcounter{figure}{0}    
\setcounter{table}{0}    

\end{document}